\pdfoutput=1

\documentclass[twocolumn]{aastex63}
\usepackage{natbib}
\usepackage{hyperref}
\usepackage{lineno}
\usepackage{ulem}
\usepackage{float}

\newcommand{\htwo}{H\,{\sc ii}~}

\shorttitle{Hourglass Magnetic fields in G35.20-0.74}
\shortauthors{O.~R. Jadhav et al.}

\begin{document}
\title{Investigation of Hourglass-shaped Magnetic fields in the G35.20–0.74 Star-Forming Complex}
\correspondingauthor{O.~R. Jadhav}
\email{Email: omkar@prl.res.in}

\author[0009-0001-2896-1896]{O.~R. Jadhav}
\affiliation{Astronomy \& Astrophysics Division, Physical Research Laboratory, Navrangpura, Ahmedabad 380009, India}
\affiliation{Indian Institute of Technology Gandhinagar Palaj, Gandhinagar 382355, India}

\author[0000-0001-6725-0483]{L.~K. Dewangan}
\affiliation{Astronomy \& Astrophysics Division, Physical Research Laboratory, Navrangpura, Ahmedabad 380009, India}

\author[0000-0002-7367-9355]{A.~K. Maity}
\affiliation{Astronomy \& Astrophysics Division, Physical Research Laboratory, Navrangpura, Ahmedabad 380009, India}

\author[0000-0002-7125-7685]{Sanhueza, Patricio}
\affiliation{Department of Astronomy, School of Science, The University of Tokyo, 7-3-1 Hongo, Bunkyo, Tokyo 113-0033, Japan}

\author[0000-0001-9312-3816]{D.~K. Ojha}
\affiliation{Tata Institute of Fundamental Research (TIFR), Homi Bhabha Road, Colaba, Mumbai - 400005, India}

\author[0000-0001-5731-3057]{Saurabh Sharma}
\affiliation{Aryabhatta Research Institute of Observational Sciences, Manora Peak, Nainital 263002, India}

\author[0000-0002-6740-7425]{Ram Kesh Yadav}
\affiliation{National Astronomical Research Institute of Thailand (Public Organization), 260 Moo 4, T. Donkaew, A. Maerim, Chiangmai 50180, Thailand}

\author[0000-0003-4941-5154]{A. Haj Ismail}
\affiliation{College of Humanities and Sciences, Ajman University, 346 Ajman, United Arab Emirates}

\author[0000-0002-2726-1388]{Moustafa Salouci}
\affiliation{Strategic Planning and Institutional Identity Administration, King Faisal University, Al-Ahsa, Saudi Arabia}

\author[0000-0003-3017-4418]{Ian Stephens}
\affiliation{Department of Earth, Environment, and Physics, Worcester State University, Worcester, MA01602, USA}
\affiliation{ Center for Astrophysics | Harvard and Smithsonian, 60 Garden Street, Cambridge MA 02138 }

\begin{abstract}

To investigate the role of magnetic fields toward the G35N and G35S sub-regions in the G35.20-0.74 star-forming complex, we utilized multi-wavelength polarimetric observations from the SOFIA/HAWC+ at 154 $\mu$m and ACT at 220 GHz/1.3 mm. The ACT 220 GHz polarization data (resolution $\sim$1$'$) show an hourglass-shaped plane-of-sky magnetic field morphologies toward both the sub-regions, although with distinct symmetry axes. SOFIA/HAWC+ 154 $\mu$m data (resolution $\sim$13\rlap.{$''$}6) confirm an hourglass morphology in G35N, whereas G35S displays a different magnetic field configuration compared to the ACT observations. 
An hourglass morphology identified at clump scales ($\sim$pc) toward G35N is consistent with the previously reported B-field morphology at core scales ($\sim$0.05 pc), supporting the scenario of a magnetically regulated collapse. 
Using the SOFIA/HAWC+ data, we estimate magnetic field strengths of $\sim$600 $\pm$ 200 $\mu$G in G35N and $\sim$850 $\pm$ 310 $\mu$G in G35S. Energy balance analysis suggests that gravity and magnetic fields contribute comparably in G35N, while in G35S the gas dynamics are dominated by magnetic field, followed by gravity and turbulence. The higher field strength in G35S likely results from compression by the expanding \htwo region, highlighting the impact of stellar feedback. 
The derived magnetic field strengths and corresponding magnetic energies should be treated as upper limits due to unresolved beam-scale correlations and the limited fitting range of the polarization angle structure function.
%
%
Overall, our results show that magnetic fields decisively regulate star formation, with G35N shaped by magnetically controlled collapse and G35S being strongly influenced by stellar feedback.
	
\end{abstract}


\section{Introduction}

Massive stars ($\gtrsim$8 $M_\odot$) profoundly influence their surroundings through intense radiative and mechanical feedback \citep{Churchwell_2006}. Yet, their exact formation mechanisms remain poorly understood \citep{Motte_2018}. In recent years, elongated structures of gas and dust, known as filaments, have attracted significant attention, as they are closely linked to star formation activity \citep{Arzoumanian_2011,Andre_2014}. These filaments organize themselves into hub-filament systems (HFSs), where multiple parsec-scale filaments converge onto a central dense hub. These hubs are characterized by high column densities ($>10^{22}$ cm$^{-2}$) and serve as key sites of mass accumulation and massive star formation \citep{Trevino_2019}. 
However, once the massive stars form and begin to influence their surroundings through strong radiation pressure and ionizing feedback, they rapidly erase the initial conditions that led to their formation. Hence, it is essential to study massive star-forming regions in the early stages, where the impact of stellar feedback is still minimal \citep{Sanhueza_2012, Sanhueza_2019,Morri_2023}. At the same time, investigations of relatively evolved regions are equally important to assess the feedback of massive stars on their surroundings.
In such environments, dust grains are believed to align with the local magnetic field (B-field) through the B-Radiative Torque Alignment (B-RATs) mechanism \citep{Lazarian_2007, Andersson_2015, Hoang_2016}. In this process, anisotropic radiation fields exert torques on irregularly shaped dust grains, spinning them up and causing their short axes to align preferentially parallel to the B-field lines. As a result, the thermal emission from these aligned dust grains becomes polarized perpendicular to the B-field direction, making it an effective tracer of the plane-of-sky (POS) B-field morphology in star-forming regions. In the case of optically thin dust emission, the POS B-field orientation can be inferred by rotating the polarization vectors by 90$^\circ$ \citep{Andersson_2015}. Alternatively, other alignment mechanisms, such as k-Radiative Torque Alignment (k-RATs), have also been proposed to explain polarized dust emission; however, observational evidence supporting this mechanism remains very limited \citep{Hull_2022,Le_2023}. In this paper, we consider the B-RATs mechanism as the dominant process responsible for dust grain alignment.

Recent dust polarization observations have revealed ordered POS B-field morphologies toward filaments and HFSs \citep{Soler_2013,Soler_2017,Planck_2016,Arzoumanian_2021}. Both observational and numerical studies suggest that B-fields can regulate gas dynamics by channeling accretion flows toward the hubs along the filaments, controlling fragmentation, and moderating the collapse of dense clumps \citep[e.g.,][]{Zhang_2014,Beuther_2018,Wang_2019, Cortes_2021,Wang_2022,Hwang_2022,Wang_2024,Khan_2024}. 
Despite this growing evidence, the precise role of B-fields in driving massive star formation, and their interplay with gravity and turbulence remains poorly understood. 

In this context, the present paper focuses on the G35.20-0.74 (hereafter G35 cloud) located at a distance of 2.19 kpc \citep[e.g.,][]{Dewangan_2017}, which is a star-forming complex, active in both low- and high-mass star formation \citep{Birks_2006,Zhang_2009,Paron_2010}. The G35 cloud hosts two major star-forming clumps/sub-regions; G35.20N (G35N hereafter) and G35.20S (G35S hereafter). G35N hosts infrared (IR)-dark clouds that provide an opportunity to probe the initial conditions of star formation, while G35S is an evolved star-forming region. Together, they form an excellent laboratory to study the role of B-fields in both early and evolved stages of massive star formation at clump scales. The distance of 2.19 kpc is adopted for the present work. 

G35N is associated with massive star formation activities, and has been extensively explored using multi-wavelength studies including Atacama Large Millimeter/sub-millimeter Array (ALMA) \citep[e.g.,][]{Monge_2014, Beltran_2016, Dewangan_2017}. \citet{Monge_2014} revealed elongated filamentary structure harboring massive dense cores toward G35N using the ALMA 850 $\mu$m data \citep[also see][]{Beltran_2016}. More recently, \citet{Hwang_2025} investigated the B-field morphology in this region using the ALMA 1.3 mm polarization data at $\sim$0.01 pc resolution. Their study revealed that the most massive core in the structure is magnetically supercritical, whereas the surrounding filament remains subcritical. The G35S clump hosts an \htwo region (i.e., Ap 2-1 Nebula) powered by an early B-type star \citep{Paron_2010} and a intermediate-mass cluster, Mercer-14 \citep{Froebrich_2011}. However, these studies were focused only on the sub-regions G35N and G35S in the G35 cloud. \citet{Dewangan_2017} explored the gas dynamics and physical process of the entire molecular cloud associated with the star-forming complex G35.20-0.74 using the $^{13}$CO(1--0) molecular line data from Galactic Ring Survey (GRS). The study suggested that cloud-cloud collision (CCC) may have caused the massive star formation activity in the G35 cloud.

Despite these extensive studies, the influence of B-fields and their interplay with turbulence and gravity in shaping star formation toward G35N and G35S remains poorly constrained. While previous studies have focused either on individual clumps or on gas kinematics, none have combined multi-wavelength polarimetry to investigate the B-field structure of G35N and G35S across large to small scales. A detailed investigation is therefore necessary to determine whether B-fields act as a dominant stabilizing force or play only a secondary role relative to turbulence and gravity in G35N and G35S. 
In this work, we try to address these gaps by examining the B-field morphology with multi-wavelength polarization datasets, estimating the field strength, and evaluating the energy balance in the star-forming clumps G35N and G35S.
In addition, multi-wavelength datasets ranging from the near-Infrared (NIR) to radio have been used to study the distribution of dust, ionized emission, and gas kinematics.

This paper is organized as follows. In Section~\ref{sec:data}, we describe the data selection. In Section~\ref{sec:results}, we present the outcomes of our multi-wavelength data analysis. In Section~\ref{sec:disc}, we discuss the implications of our results. Finally, the findings are summarized and the conclusions are presented in Section~\ref{sec:summary}.

\section{Datasets}
\label{sec:data}

The goal of this paper is to investigate the role of the B-fields and to study the overall physical environment of the G35 cloud. This requires a combination of multi-wavelength observations. We used NIR and mid-infrared (MIR) data to trace the warm dust emission and study absorption features associated with the cloud. To examine the B-field morphology, we employ the 154 $\mu$m far-infared (FIR) polarization data. At larger scales, the B-field morphology is studied using the {\it Planck} 353 GHz polarization measurements. To estimate the B-field strength and study the gas kinematics, we incorporate millimeter $^{13}$CO(1--0) molecular line data along with the 220 GHz polarization maps. Finally, to study the ionized emission toward the target region, we have used the 1.3 GHz radio continuum observations. A brief description of these datasets is provided below.

\subsection{NIR and MIR data}

We acquired NIR and MIR images at 3.6--8.0 $\mu$m (resolution $\sim$2$''$; plate scale $\sim$0\rlap.{$''$}6) from the {\it Spitzer} Galactic Legacy Infrared Mid-Plane Survey Extraordinaire \citep[GLIMPSE;][]{Benjamin_2003} survey. 
The MIR 12.0 $\mu$m image was collected from the Unblurred Coadds of the Wide-field Infrared Survey Explorer (WISE) Imaging (unWISE) survey \citep{Lang_2014}, which provides WISE images with an improved signal-to-noise ratio. The improved signal-to-noise ratio allows the unWISE 12.0 $\mu$m image to trace large-scale absorption features more effectively than the {\it Spitzer} 8.0 $\mu$m IRAC data.

\subsection{FIR data}
The G35N and G35S sub-regions were observed with the Stratospheric Observatory for Infrared Astronomy (SOFIA) telescope using the High-resolution Airborne Wideband Camera Plus (HAWC+) instrument \citep{Harper_2018} at 154 $\mu$m (or Band D). 
These polarization observations were taken as part of a survey program (PI: Ian Stephens; Proposal ID: 09$\_$0016) and were obtained in band D (154 $\mu$m) using the on-the-fly scan mapping mode with a Lissajous scan pattern. The telescope scanned the target with a commanded scan speed of 200$''$ s$^{-1}$ and amplitudes of 111$''$ and 138$''$ in elevation and cross-elevation, respectively. Three scan iterations were performed. The total on-source integration time was 2481 seconds. 
The data were reduced using the standard HAWC+ data reduction pipeline (version 3.2.0), which performs flat-fielding, despiking, instrumental polarization correction, and iterative removal of correlated atmospheric and instrumental noise using a maximum-likelihood map-making algorithm. The reduction employed the standard configuration including the neighbour-based despiking algorithm (level = 10), motion and whitening filters, and correlated-noise removal from the sky signal, detector rows, multiplexer channels, and polarization arrays.	
In this study we have used the  debiased HAWC+ polarization data delivered as part of the Level 4 products, which have a spatial resolution of $\sim$13\rlap.{$''$}6 and a pixel scale of $\sim$3\rlap.{$''$}4.

\subsection{Sub-millimeter data}

To study the cold dust emission, the 870 $\mu$m emission data (resolution $\sim$18\rlap.{$''$}2)  were sourced from the Atacama Pathfinder Experiment (APEX) Telescope Large Area Survey of the GALaxy  \citep[ATLASGAL;][]{Schuller_2009}. 
The catalog by \citet{Urquhart_2017} details of about 8000 dense clumps in the Galactic disc using the ATLASGAL data. They also provided the velocities, distances, luminosities, and masses of clumps \citep[see][for more details]{Urquhart_2017}. Using this catalog, we identified the positions of dense clumps associated with the G35 cloud.
To infer the POS B-field orientation toward the cloud at larger scales ($\sim$10 pc) we utilized the 353 GHz polarization maps obtained from the {\it Planck} Legacy archive observed as part of the {\it Planck} mission. The data provides the $I$, $Q$, and $U$ polarization maps observed using the High Frequency Instrument (HFI) \citep{Lamarre_2015} on the {\it Planck} satellite, with a spatial resolution of 5$'$ \citep{Planck_2015}.

\subsection{Millimeter data}

In order to trace molecular gas associated with the selected target, the GRS $^{13}$CO(1--0) line data were employed. The GRS line data have a velocity resolution of 0.21 km s$^{-1}$, an angular resolution of 45$''$ with 22$''$ sampling, a main beam efficiency ($\eta_{mb}$) of $\sim$0.48, a velocity coverage of $\sim$5 to 135 km s$^{-1}$, and a typical rms sensitivity ($\sigma$) of $\sim$0.13 K \citep{Jackson_2006}. The antenna temperature is converted into main beam temperature using the formula $T_{\rm mb}$ = $T_{\rm A}$/$\eta_{mb}$. We used the Atacama Cosmology Telescope (ACT) 220 GHz (or 1.3 mm) DR6 polarization data to investigate the POS B-field structure toward the G35 cloud. The Stokes $I$, $Q$, and $U$ maps were obtained from the NASA Legacy Archive for Microwave Background Data Analysis (LAMBDA). These data have a spatial resolution of $\sim$1$'$ at 220 GHz \citep{Thornton_2016}.


\subsection{Radio data}
We used the South African Radio Astronomy Observatory (SARAO) MeerKAT Galactic Plane Survey (SMGPS) 1.3 GHz/23 cm continuum emission data \citep[resolution $\sim$8$''$; sensitivity $\sim$22 $\mu$Jy beam$^{-1}$;][]{Goedhart_2024} to study the ionized emission toward the target.

\section{Results}\label{sec:results}

\subsection{Physical environment of G35 cloud}
\label{phy_env}
\begin{figure*}
	\centering
	\includegraphics[width= 12 cm]{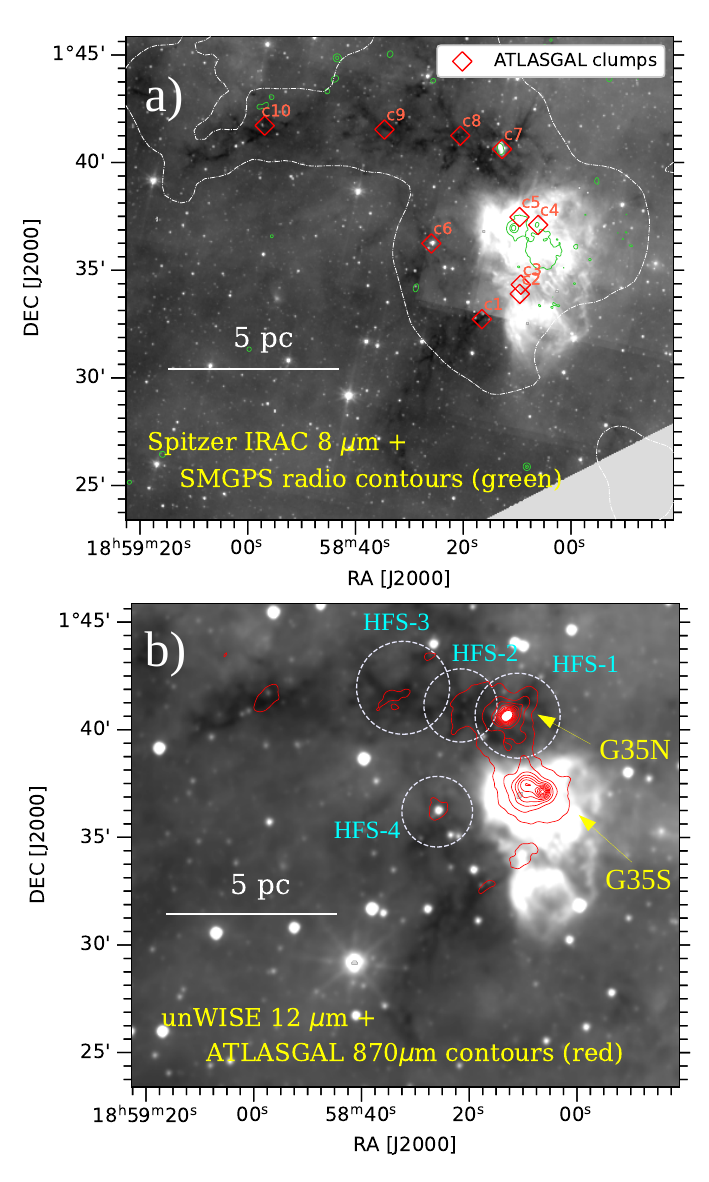}
	\caption{a) {\it Spitzer}/IRAC 8.0 $\mu$m image (in the inverse hyperbolic sine scale) overlaid with the positions of the ATLASGAL clumps at 870 $\mu$m (c1-c10) (see red diamonds). The green contours represent the SMGPS 1.3 GHz continuum emission.  The contour levels are 1, 17, 34, and 50 mJy beam$^{-1}$. The white dot–dashed contour shows GRS $^{13}$CO (1–0) integrated intensity ([30, 40] km s$^{-1}$) with a level of 13 K km s$^{-1}$ b) The panel shows the unWISE 12 $\mu$m image (in the inverse hyperbolic sine scale) overlaid with the ATLASGAL 870 $\mu$m contours (in red). The contour levels are 1, 2, 4, 6, 7, 9, 11, and 13 mJy beam$^{-1}$. The white dashed circles show the positions of the four HFS candidates detected toward the G35 cloud. 
	The scale bar corresponding to 5 pc at a distance of 2.19 kpc is shown. }
	\label{fg1}
\end{figure*}

\begin{figure*}
	\centering
	\includegraphics[width= 15 cm]{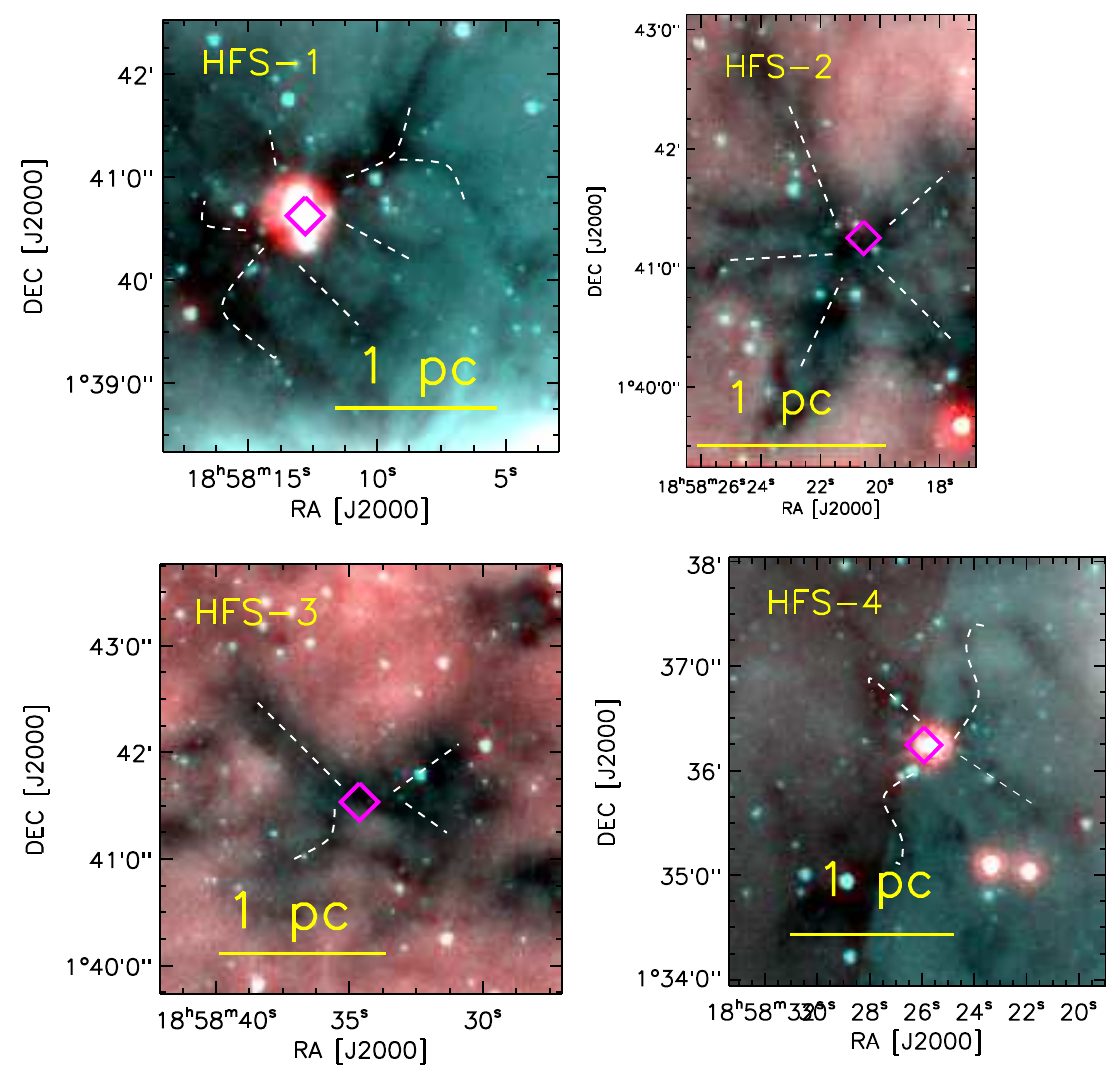}
	\caption{The panels show the two-color composite images (red: unWISE 12.0 $\mu$m; turqoise: {\it Spitzer} 8.0 $\mu$m) of the HFS candidates detected toward the G35 cloud (see dashed circles in Figure~\ref{fg1}b). The white dotted curves/lines highlights the visually identified filaments. The diamonds show the locations of the ATLASGAL clumps. The scale bar corresponding to 1 pc at a distance of 2.19 kpc is shown in each panel .}
	\label{fg1.1}
\end{figure*}

\begin{figure*}
	\centering
	\includegraphics[width= 19 cm]{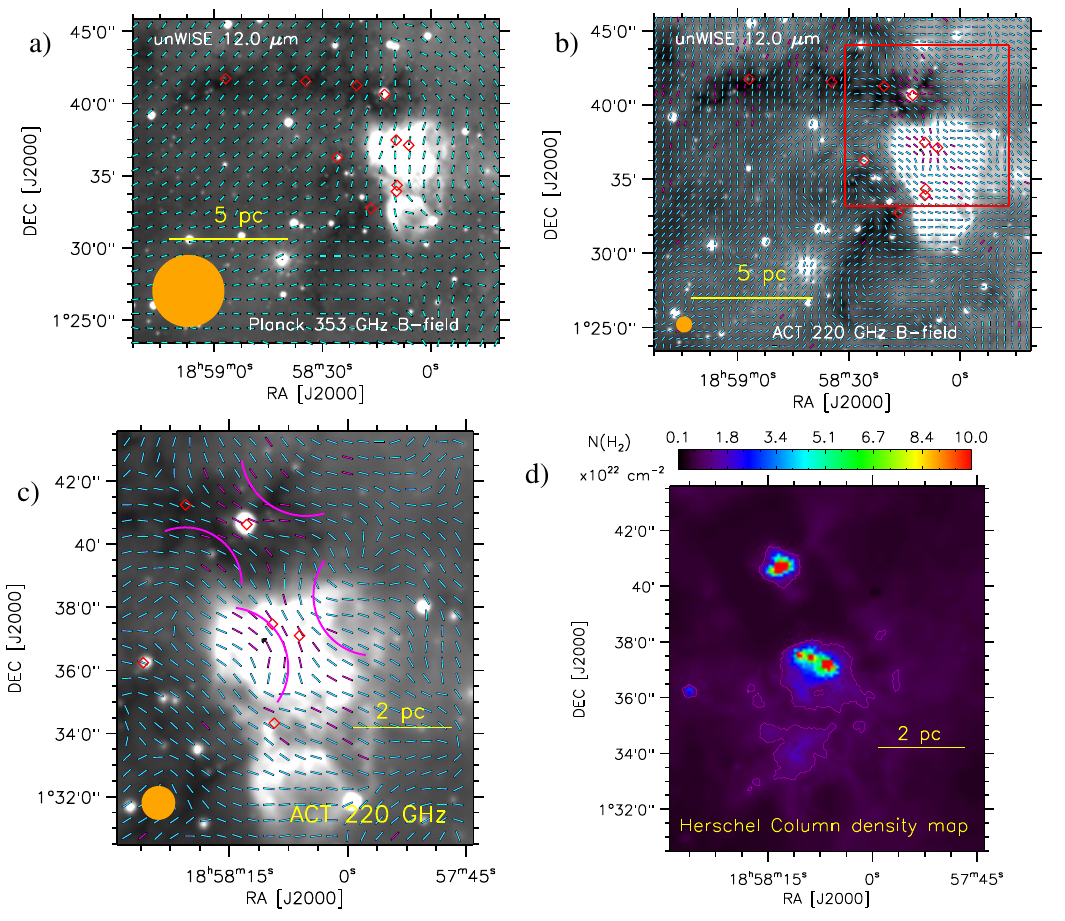}
	\caption{a) Overlay of segments (in cyan) showing the B-field orientations inferred using the {\it Planck} 353 GHz polarization data (resolution $\sim$5$'$) on the unWISE 12.0 $\mu$m map (in the inverse hyperbolic sine scale). b) unWISE 12.0 $\mu$m image (in the inverse hyperbolic sine scale) overlaid with the segments displaying the B-field orientations from the ACT 220 GHz data (resolution $\sim$1$'$). The cyan and magenta segments represent polarization vectors with $p/\sigma_p<3$ and $p/\sigma_p\geq3$, respectively. The red and yellow dot dashed box encompasses the area shown in Figure~\ref{fg3} and ~\ref{fg2}c  respectively. c) Same as Figure~\ref{fg2}b, but for the FOV indicated by the yellow box. The magenta curves highlight the hourglass-like shape of the B-fields. For display purposes, B-field segments are plotted at the pixel scale, i.e., oversampled relative to the beam resolution. d) Herschel $N(\rm H_2)$ map from PPMAP toward the area shown by yellow box in Figure~\ref{fg2}b overlaid with the $N(\rm H_2)$ contour level at 9.5 $\times$ 10$^{21}$ cm$^{-2}$. The diamonds (in red) in both panels highlight the location of the ATLASGAL clumps. The polarization data beam size is indicated by a circle placed in the bottom-left corner of the image. The scale bar derived at a distance of 2.19 kpc is shown in each panel.}
	\label{fg2}
\end{figure*}

\begin{figure*}
	\centering
	\includegraphics[width= 11 cm]{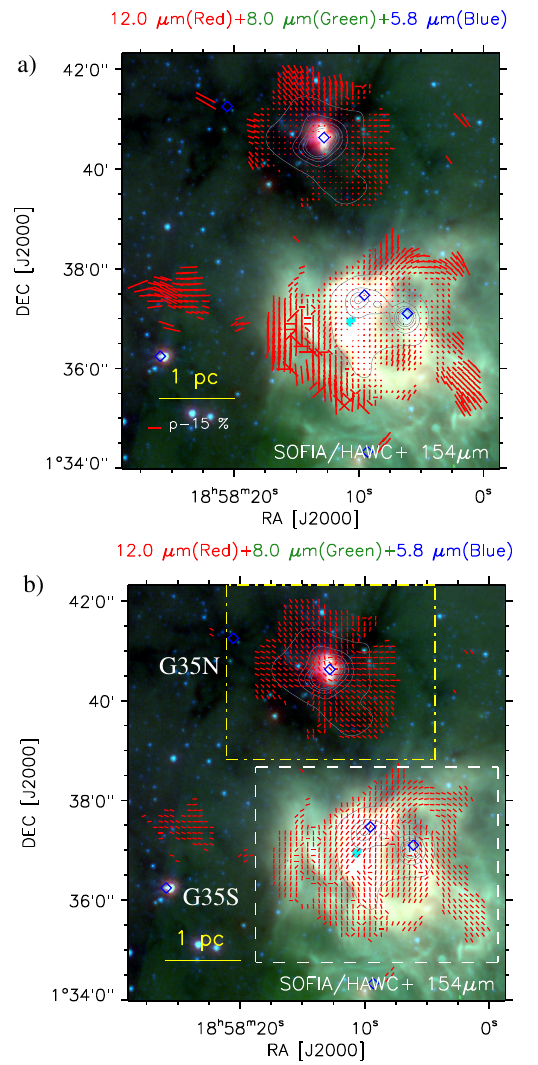}
	\caption{a) B-field segments (in red) inferred from the SOFIA/HAWC+ 154 $\mu$m data overlaid on the three-color composite image (red: 12.0 $\mu$m, green: 8.0 $\mu$m, blue: 5.8 $\mu$m). The length of the segments is proportional to the degree of polarization. A reference vector corresponding to 15$\%$ polarization is shown in the bottom-left corner of the panel. b) The panel is the same as Figure~\ref{fg3}a, except that the segment lengths are kept constant. The yelow dot-dashed box displays the field shown in figure~\ref{fg4}a. The white dashed box highlights the G35S region. The grey contours represent the SOFIA/HAWC+ 154 $\mu$m continuum emission. The contour levels are 1, 3.8, 6.5, 9.3, and 12 Jy beam$^{-1}$. The diamonds show the locations of the ATLASGAL clumps. In each panel, the scale bar corresponding to 1 pc at a distance of 2.19 kpc is shown.}
	\label{fg3}
\end{figure*}

{In this section, we examined  the spatial distribution of MIR emission, dense clumps, and ionized emission toward the G35 cloud (including the G35N and G35S sub-regions), allowing us to probe its physical environment.

Figure~\ref{fg1}a shows the {\it Spitzer} 8.0 $\mu$m image of the G35 cloud, appearing as an elongated horseshoe--like structure in absorption, with a very bright center. A total of 10 ATLASGAL clumps at 870 $\mu$m (c1-c10) are distributed toward the cloud, which are located at a distance of $\sim$2.19 kpc \citep{Urquhart_2017}. Among these, only six clumps c2--c7 are associated with the MIR emission, while the other clumps are MIR quiet. The physical properties of these clumps are also obtained from the ATLASGAL clump catalog provided by \citet{Urquhart_2017} (see Table~\ref{tab2}). The radial velocities ($V_{\rm lsr}$), dust temperatures ($T_{\rm d}$), and effective radii of the clumps are [31.8, 37.5] km s$^{-1}$, [11.8, 21.7] K, and [0.1, 1.6] pc, respectively. 
%
%

To examine the distribution of the ionized emission, we overlaid the SMGPS 1.3 GHz continuum emission contours on the {\it Spitzer} 8.0 $\mu$m image (see Figure~\ref{fg1}a). According to \citet{Dewangan_2017}, the NVSS 1.4 GHz continuum emission is exclusively detected toward the clump c5 (in G35S) only. 
However, Figure~\ref{fg1}a shows the SMGPS 1.3 GHz continuum emission associated with clumps c5, c6, and c7. 
The radio emission toward the most massive clump c7/G35N ($\sim$3100 $M_{\odot}$) is very compact, whereas the clumps c5 and c6 in G35S are associated with an extended radio emission. In Figure~\ref{fg1}b, we present the unWISE 12.0 $\mu$m image overlaid with the ATLASGAL 870 $\mu$m continuum emission contours.
%

%
High resolution {\it Spitzer} 8.0 $\mu$m and unWISE 12.0 $\mu$m images also allow to examine the substructures (in absorption) features toward the G35 cloud. 
In some areas, these MIR images show compact central regions, which seem to be surrounded by at least three dark features with high aspect ratio. In particular, such configurations are seen toward four ATLASGAL clumps, hinting at the presence of four HFS-like morphologies (HFS1--HFS4; extent $\sim$2--3 pc; see also \citealt{dewangan25n}). HFS1, HFS2, HFS3, and HFS4 are seen toward the clumps C7/G35N, C8, C9, and C6, respectively. Using the {\it Spitzer} 8.0 $\mu$m images, a zoomed-in view of each HFS candidate is shown in Figure~\ref{fg1.1}. Interestingly, two HFS candidates (i.e., HFS-1/G35N and HFS-4) are associated with IR-bright features at their centers, while the hubs of the other two HFSs are found with IR-dark emisison.
%
%
%

\subsection{Magnetic field morphology toward the G35 cloud}
\label{sec:allmag}
Using the {\it Planck}, SOFIA, and ACT polarization data,  the following subsections investigate the POS B-field morphology toward the G35N and G35S sub-regions, which are previously known active star-forming sites. 

\subsubsection{{\it Planck}  B-field morphology}
\label{sec:Planck}
To infer the POS B-field morphology at larger spatial scales ($\sim$10 pc), we used the {\it Planck} 353 GHz polarization observations \citep[resolution $\sim$5$'$;][]{Planck_2015}. The coarse beamsize of the {\it Planck} 
data allows us to study the global B-field morphology toward the G35 cloud. The $I$, $Q$, and $U$ Stokes maps are obtained from the {\it Planck} archive. The polarization angle is estimated using 
\begin{eqnarray}
\theta=\frac{1}{2}~{\rm arctan2}(-U, Q)
\end{eqnarray}
The minus sign is used to convert the polarization angles to IAU convention. We obtained the B-field position angles by rotating the polarization vectors by 90$^\circ$. 
Figure~\ref{fg2}a shows an overlay of the segments showing the {\it Planck} 353 GHz B-field orientations on the unWISE 12.0 $\mu$m image. The field is oriented along the north-south direction in the G35N and G35S sub-regions, whereas it is oriented along the east-west direction in the southern IR-dark region of the G35 cloud.

\subsubsection{ACT B-field morphology}
\label{sec:act}
To further investigate the B-field morphology at smaller spatial scales ($\sim$5 pc), we use polarization observations from the ACT at 220 GHz (resolution $\sim$1$'$). Previously, \citet{Guan_2021} used the ACT polarization data to derive the polarization maps toward the galactic center, which remains the only study in our galaxy using this dataset. The polarization angles are computed in the same manner as described in Section~\ref{sec:Planck}, except that $U$ is used instead of $-U$, since the ACT polarization angles already follow the IAU convention. To increase the SNR, the polarization angle map is generated using the $Q$ and $U$ maps, which were smoothed using a Gaussian kernel with 1 pixel ($\sim$30$''$) standard deviation, yielding a final resolution of $\sim$1\rlap.{$'$}07. The resulting map of B-field orientations is shown in Figure~\ref{fg2}b. 
The ACT data enable us to trace much finer B-field structures toward the cloud compared to the {\it Planck} polarization data. Figure~\ref{fg2}c show the B-field orientations from the ACT 220 GHz toward the G35N and G35S sub-regions. In the direction of these sub-regions, hourglass-shaped morphologies seem to be evident; however, the axes of symmetry for the two hourglass shapes differ from each other. 
Due to the coarse beam size of the {\it Planck} data ($\sim$5$'$), these structures are not visible in the {\it Planck} 353 GHz B-field map (see Figure~\ref{fg2}a). 


\subsubsection{SOFIA/HAWC+ B-field morphology}

To explore the POS B-field morphology at ($\sim$pc scale), we use the polarization observations from the SOFIA/HAWC+ at 154 $\mu$m (band D), toward G35N and G35S. The polarization angles are calculated in the same way as in Section~\ref*{sec:act}. Figure~\ref{fg3}a shows the overlay of the segments showing the B-field orientations inferred from the SOFIA/HAWC+ 154 $\mu$m data on the IR three color composite image (red: 12.0 $\mu$m; green: 8.0 $\mu$m; blue: 5.8 $\mu$m).  We only retained the vectors which satisfy the following criteria: $p/\sigma_p>$3, $I/\sigma_I>$10, and $p<30$ \citep[same as][]{Ngoc_2023}.

After applying these selection criteria, more than 97$\%$ of the retained vectors lie above the 0.3--0.5$\%$ polarization fraction range, where instrumental polarization dominates \citep{Harper_2018}. Additionally, the effective SNR of the retained B-field vectors is much higher than these quality cuts.

In Figure~\ref{fg3}a, the length of the segments represents the degree of polarization ($p$). The degree of polarization is higher in low-intensity regions and decreases toward areas of higher intensity (see gray contours in Figure~\ref{fg3}a). This behavior is observed in both the G35N and G35S clumps. Figure~\ref{fg3}b shows the same B-field orientations, where the length of the segments is kept constant to better understand the B-field morphology.  The field lines are seen to radially aligned toward the center of the HFS-1/G35N, with the overall morphology resembling to an hourglass. However, the B-field morphology toward G35S appears somewhat random, yet maintains an overall organized structure.
When examined carefully, the G35N clump retains the hourglass-shaped morphology seen in the ACT 220 GHz data, exhibiting the same axis of symmetry. In contrast, the G35S clump does not preserve the  field morphology from the ACT data, but instead closely resembles the B-field structure observed in the {\it Planck} 353 GHz data. The implication of this result discussed in Section~\ref{sec:extinction polarization}.

Figure~\ref{fg4}a presents a zoomed-in view of HFS-1/G35N, overlaid with B-field streamlines from the SOFIA/HAWC+ data. The streamlines clearly trace an hourglass-shaped pattern, with the hourglass axis passing through the hub of the HFS.
Figure~\ref{fg4}b presents the polar histogram of the B-field position angles toward G35N (orange), and G35S (blue). The peaks at 0$^o$ and 180$^o$ mainly arise from the position angles in the G35S region, whereas the peak at 30$^o$ is primarily contributed by G35N. Because polarization position angles are defined modulo 180$^o$, the apparent peaks at 0$^o$ and 180$^o$ correspond to the same orientation. Thus, the G35S distribution effectively reflects a single preferred orientation, while G35N shows a more uniform, nearly symmetric spread between 0$^o$ and 90$^o$.

\subsubsection{Comparison of {\it Planck}, ACT, and SOFIA polarization data}
We convolved the ACT 220 GHz observations to match the angular resolution ($\sim$5$'$) and pixel scale ($\sim$1$'$) of the {\it Planck} 353 GHz polarization data (see Figure \ref{fig:afg3} in Appendix). The B-field orientations inferred from the {\it Planck} and ACT datasets are consistent toward the northern and southern IR-dark filament. However, the B-field orientations toward the G35N and G35S sub-regions from both datasets differ. This discrepancy arises because \textit{Planck} retains the large-scale polarization signal (cloud-scale and Galactic-scale field morphology), while ACT, as a ground-based instrument, filters out these large-scale modes during spatial filtering.

To investigate these differences in more detail, we examined whether the small-scale B-field morphology observed in the ACT data toward G35N and G35S is also evident in the SOFIA/HAWC+ observations. We find that the hourglass-shaped B-field morphology toward G35N, evident in the ACT 220 GHz data, is also retained in the SOFIA/HAWC+ 154 $\mu$m observations. In contrast, no such morphology is observed toward G35S. This raises the question of whether the SOFIA data can reliably trace B-field morphology in IR-dark clouds. To address this, we convolved the $Q$ and $U$ maps from the SOFIA/HAWC+ Band D observations to match the angular resolution and pixel scale of ACT (resolution $\sim$1\rlap.{$'$}1; pixel scale $\sim$30$''$), and overlaid the resulting SOFIA/HAWC+ and ACT B-field vectors. Figure~\ref{fg7}a shows this comparison, with B-field orientations from the SOFIA/HAWC+ Band D in red and from the ACT 220 GHz in cyan. The two datasets exhibit a close alignment in G35N, whereas toward G35S, their field orientations differ.  The implication of this result is discussed in Section~\ref{sec:extinction polarization}.

\subsection{Magnetic field strength}
\label{B-field strength}

The high-resolution SOFIA/HAWC+ 154 $\mu$m polarization observations toward the G35 allow us to estimate the POS B-field strength ($B_{\rm POS}$) toward the G35N and G35S clumps. The $B_{\rm POS}$ value is estimated using the Davis-Chandrasekhar-Fermi (DCF) method \citep{Chandrasekhar_1953}. However, we use a variant of the DCF method by \citet{Crutcher_2012} to estimate the $B_{\rm POS}$ using, 
\begin{equation}\label{eq:dcf}
B_{\rm POS} \approx 9.3~\sqrt{n(\rm H_2)}~\frac{\Delta V_{\rm nt}}{\delta_{\theta}}~(\mu G),
\end{equation}
where $n(\rm H_2)$ is the volume density in cm$^{-3}$, $\Delta V_{\rm nt}$ is the FWHM of the non-thermal velocity component in km s$^{-1}$, and $\delta_{\theta}$ is the polarization angle dispersion in degrees. In the following subsections, we derive these parameters for both the G35N and G35S sub-regions to estimate $B_{\rm POS}$ values.

\subsubsection{Volume density}

Assuming a spherical morphology, the volume density of a clump can be calculated as follows,
\begin{eqnarray}
n({\rm H_2}) = \frac{3M}{4\pi R^3},
\label{eqn3}
\end{eqnarray}
where $M$ is the mass, and $R$ is the effective radius of the clump which is given by R = $\sqrt{A_{\rm eff}/\pi}$, where $A_{\rm eff}$ is effective area of the clump. We used the $N(H_2)$ map (Figure~\ref{fg3}d) generated by \citet{Marsh_2017} using the point process mapping (PPMAP) algorithm  to calculate the mass of the G35N and G35S clumps. The clumps were identified using the {\it astrodendro} package to estimate the mass. We used the following equation \citep[e.g.,][]{Dewangan_2017}:
\begin{equation}
M_{\rm clump}= \mu_{\rm H_{2}}~m_{\rm H}~ a_{\rm pixel}~\Sigma N(\mathrm H_2),
\label{eq1}
\end{equation}
 where $\mu_{\rm H_{2}}$ is the  ${\rm H_2}$ mean molecular weight (assumed to be 2.8), $m_{\rm H}$ is the mass of the hydrogen atom, $a_{\rm pixel}$ is the physical area subtended by 1 pixel in cm$^2$, and $\Sigma N(\mathrm H_2)$ is the integrated column density. The mass and radius of the G35N clump are 550 $\pm$ 280 $M_\odot$ and 0.33 pc, respectively, while the mass and radius of the G35S clump are 940 $\pm$ 470 $M_\odot$ and 0.53 pc respectively.


Substituting these values in equation~\ref{eqn3}, we obtain the $n({\rm H_2})$ toward G35N and G35S as 5.3 $\pm$ 2.6 $\times$ 10$^4$ cm$^{-3}$ and 2.1 $\pm$ 1.0 $\times$ 10$^4$ cm$^{-3}$ respectively. The uncertainty in $n({\rm H_2})$ mainly arises due to the uncertainty in the mass estimation which is $\sim$50$\%$ \citep{Sanhueza_2019}. However, this represents a lower limit; the true uncertainty in $n({\rm H_2})$ may be several times higher due to the unknown line-of-sight projection and the assumption of spherical morphology.

\subsubsection{Velocity dispersion}

We use the $^{13}$CO(1--0) molecular line data to estimate  $\Delta V_{\rm nt}$ for the G35N and G35S sub-regions. We obtained $\sigma_{\rm v}$ using the $^{13}$CO(1--0) moment-2 map. The FWHM of the velocity component is calculated as, $\Delta V$ = 2.355 $\times$ $\sigma_{\rm v}$. However, this value also consists the contribution of the thermal component of the velocity. To obtain the non-thermal velocity component we use the equation \citep{Fuller_1992, Kauffmann_2013}
\begin{equation}
{\Delta V_{\rm nt}}^2 = {\Delta V}^2 - {\rm 8~ln2}\frac{kT}{m_{^{13}CO}} ,
\end{equation}
where $k$ is the Boltzmann constant, $T$ is the gas temperature taken as 15 K (similar to the mean $T_{\rm d}$ of the ATLASGAL clumps), and $m_{^{13}CO}$ is the mass of $^{13}$CO molecule which is 29 amu. Figure~\ref{fg5} shows the $^{13}$CO(1--0) moment-2 map ($\Delta V$) toward the G35N and G35S. The mean $\Delta V_{\rm nt}$ values toward G35N and G35S are 2.98 $\pm$ 0.43 km s$^{-1}$ and 3.84 $\pm$ 0.29 km s$^{-1}$, respectively.

\subsubsection{Polarization angle dispersion}
We use the structure function method to estimate the polarization angle dispersion toward G35N and G35S using the SOFIA/HAWC+ 154 $\mu$m polarization data. \citep{Hildebrand_2009,Houde_2009}. According to this method, the $B_\mathrm{POS}$~can be estimated using Eq.~\ref{eq1} but the polarization angle dispersion $\delta\theta$ is replaced by the structure function of the polarization angles, so-called angular structure function. The angular structure function, $D^{1/2}_\theta$, at an angular scale $\ell$ is estimated as \citep{Hildebrand_2009}:
\begin{equation}\label{eq:SF}
D^{1/2}_\theta(\ell) \equiv \left< \Delta \theta^2 (\ell) \right> ^{1/2}
= \left\{ \frac{1}{N(\ell)} \sum_{i=1}^{N(\ell)} \Bigl[ \theta(x) - \theta(x+\ell)\Bigr]^2\right\}^{1/2}, 
\end{equation} 
where $\left< ... \right>$ denotes an average, $\Delta\theta(\ell) \equiv\theta(x) - \theta(x+\ell)$ is the polarization angle difference between individual pairs of polarization vectors at position $x$ and $x+\ell$, where $\theta(x)$ and $\theta(x+\ell)$ are the corresponding polarization angles at position $x$ and $x+\ell$, respectively, and $N(\ell)$ is number of pairs of vectors with a displacement of $\ell$. 
At scales $\ell$ much smaller than the scale for a variation of the large-scale B-field structure (typically of $\sim$5$'$, as seen by the {\it Planck} observations), the total angular dispersion function can be expressed as \citep{Hildebrand_2009}:
\begin{equation}\label{eq:quadratic_func}
D_\theta(\ell) \simeq b^2 + m^2\ell^2 + \sigma_\mathrm{M}^2(\ell), 
\end{equation}
where $b$ represents the turbulent component of B-fields ($B_\mathrm{t}$), $m\ell$ describes the large-scale structure of B-fields ($B$), and $\sigma_\mathrm{M}(\ell)$ is subject to measurement uncertainties. Also, the ratio between the turbulent and large-scale B-field strength indicates the angular dispersion, given as \citep{Hildebrand_2009}:
\begin{equation}
\delta\theta\simeq \frac{\left<B^2_\mathrm{t}\right>^{1/2}}{B} =\frac{b}{\sqrt{2-b^2}}.
\end{equation}
When $B_\mathrm{t} \ll B $, i.e., $b\ll1$~rad, we can obtain $\delta\theta \sim b/\sqrt{2}$. We calculated the $D^{1/2}_\theta(\ell)$ for G35N and G35S using equation~\ref{eq:SF} and fitted with the quadratic function given as equation~\ref{eq:quadratic_func}. The plots for both the sub-regions are shown in Figure~\ref{fg6}. We only fitted the data points between $\sim$13\rlap.{$''$}6 and $\sim$40\rlap{$''$}, to avoid fitting large-scale structures.
The lower limit for fitting was taken as the resolution of the polarization data, while the upper limit was determined by visual inspection of the structure function, selecting the scale just before saturation.
The best-fit parameters for G35N and G35S are b = 15.11$^o$ $\pm$ 1.39$^o$ and 6.43$^o$ $\pm$ 1.19$^o$ respectively.
The polarization angle dispersion ($\delta_{\theta}$) for the northern and southern clumps is 10.7$^o$ $\pm$ 1.7$^o$ and 6.0$^o$ $\pm$ 1.6$^o$, respectively.

Following \citet{Hwang_2021}, we also used the sliding box method to generate a pixel-by-pixel polarization angle dispersion map. We proceed by defining a box of 9 $\times$ 9 pixels and 15 $\times$ 15 pixels  (pixel scale $\sim$3\rlap.{$''$}4), which is approximately twice and three times the beam size of the SOFIA/HAWC+ band D observations ($\sim$13\rlap.{$''$}6) centered on the $i$th pixel \citep{Guerra_2021, Ngan_2024}, respectively. We estimated the root-mean-squared (RMS) of all the pixel values inside the box using the following formula: $\sigma_\theta = \sqrt{\frac{\sum_{i=1}^N (\theta_i - \bar{\theta})^2}{N}}$. Here, \(N\) is the total number of pixels within the box. This RMS value corresponds to the polarization angle dispersion at $i^{th}$ pixel. This process is repeated by sliding the box across the entire region to generate the polarization dispersion map. The dispersion value for a given box is computed only if more than 50$\%$ of the pixels within it have valid (non-NaN) values; otherwise, the central pixel is excluded from the calculation. The resulting dispersion values are debiased by subtracting the contribution from measurement noise using, $\sigma_{\mathrm{true}} = \sqrt{\sigma_{\mathrm{obs}}^{2} - \sigma_{\mathrm{noise}}^{2}}$. Figures~\ref{fg5}b and \ref{fg5}c show the resulting polarization dispersion map. The mean polarization angle dispersion values toward G35N and G35S are 9.17$^\circ$ $\pm$ 5.18$^\circ$ and 9.03$^\circ$  $\pm$ 5.03$^\circ$, respectively, using the 9 $\times$ 9 pixels box. While the angle dispersion values toward the G35N and G35S sub-regions are 11.96$^\circ$ $\pm$ 5.64$^\circ$ and 11.92$^\circ$  $\pm$ 5.60$^\circ$, respectively, using the 15 $\times$ 15 pixels box. The increase in dispersion for larger box sizes is expected \citep{Cortes_2024}, as larger windows sample broader-scale variations in the B-field; however, the values remain consistent within the errorbars between two cases.

In the direction of G35N and G35S, the derived polarization dispersion values using the structure function and the sliding box method are nearly identical, and cannot be distinguished within the error bars. For the B-field strength calculations, we therefore use the $\delta_{\theta}$ values obtained from the structure function method, as this method provides smaller uncertainties compared to the sliding box approach.
We note that the agreement between the model and the observed structure function is not perfect, although the model captures the overall trend at intermediate spatial scales ($18.2 < l < 40$). Deviations at small separations ($l < 18.2$) may arise from beam smoothing, while differences at larger separations ($l > 40$) may reflect the onset of saturation in the structure function. A more complete treatment following \citet{Houde_2009}, which explicitly accounts for beam smoothing and LOS averaging, would therefore be a valuable extension. Consequently, the magnetic field strengths derived in this study should be regarded as upper limits.}


Substituting the values of $n(\rm H_2)$, $\Delta V_{\rm nt}$, and $\delta\theta$ in equation~\ref{eq:dcf}, we obtain the $B_{\rm POS}$ values toward the G35N and G35S sub-regions as 
$\sim$600 $\pm$ 200 $\mu$G and $\sim$850 $\pm$ 310 $\mu$G,  respectively. However, the uncertainity in $B_{\rm POS}$ can reach up to a factor of a few owing to the large uncertainties associated with the $n(\rm H_2)$ values.

\subsubsection{Mass-to-flux ratio}

To infer the relative importance of B-fields with respect to gravity, we use a parameter known as the mass-to-flux ratio, $\lambda$, which is estimated using the relation given by \citep{Crutcher_2004}

\begin{equation}
\lambda_{\rm obs}=\frac{(M/\Phi)_{\rm observed}}{(M/\Phi)_{\rm critical}}=7.6\times10^{-21}\frac{N(\rm H_2)}{B_{\rm tot}},
\end{equation}
where $B_{\rm tot}$ is the total B-field strength in $\mu$G which is given as, $B_{\rm tot}$ = 1.3 $\times$ $B_{\rm POS}$ \citep{Crutcher_2004}. If $\lambda>1$ signifies the dominance of gravity over the B-field, hence referred to as magnetically supercritical. Conversely, if $\lambda<1$, it denotes that the B-field is dominant over gravity, characterizing those areas as magnetically subcritical. The mass-to-flux ratio toward G35N and G35S is 0.7 $\pm$ 0.5 and 0.3 $\pm$ 0.2 respectively. These values indicate that G35N and G35S clumps are in a magnetically subcritical state, while G35N could be in a transcritical state within error limits.

\subsection{Energy Balance}
\label{sec:energy}
Star formation activity in a cloud is driven by a complex interplay of gravity, turbulence, and B-fields. Therefore, quantifying the relative contributions of these forces is essential to understand their roles in the physical processes governing star formation. \citet{Chung_2022} present the virial theorem including these energies.
\begin{eqnarray}
\frac{1}{2} \text{\"{I}}~=2E_{\rm K} + E_{\rm B} + E_{\rm G}
\end{eqnarray}
The gravitational energy for a spherical source is given by \citep{Fiege_2000}

\begin{equation}\label{Eg}
E_{\rm G} = \frac{3GM^2}{5R}.
\end{equation}
The kinetic energy for a sphere is given by \citep{Fiege_2000}
\begin{eqnarray}\label{Ek}
E_{\rm K} = \frac{3}{2}M\sigma_V^2.
\end{eqnarray}
The magnetic energy is given by \citep{Sanhueza_2025}
\begin{eqnarray}\label{Eb}
E_{\rm B} = \frac{1}{6}B^2R^3.
\end{eqnarray}

The $E_{\rm G}$, $E_{\rm K}$, and $E_{\rm B}$ values for G35N are 4.73 $\pm$ 4.73 $\times$ 10$^{46}$ ergs, 2.66 $\pm$ 1.52 $\times$ 10$^{46}$ ergs, and 9.80 $\pm$ 4.60 $\times$ 10$^{46}$ ergs, respectively. For G35S, the $E_{\rm G}$, $E_{\rm K}$, and $E_{\rm B}$  values are 8.42 $\pm$ 8.42 $\times$ 10$^{46}$ ergs, 7.49 $\pm$ 3.89 $\times$ 10$^{46}$ ergs,  and 8.73 $\pm$ 5.06 $\times$ 10$^{47}$ ergs. 
%
In G35N, the gravitational and magnetic energies are roughly comparable, whereas in G35S, the magnetic energy appears to dominate over both the gravitational and turbulent energies.

\section{Discussion}
\label{sec:disc}
\subsection{HFSs and massive star formation}
HFSs are structures in which parsec-scale filaments converge onto a central hub characterized by high column densities ($>$10$^{22}$ cm$^{-2}$) \citep{Myers_2009}. These hubs are known to be active sites for star formation, especially massive stars \citep{Motte_2018,Kumar_2020,Zhou_2022}. However, most studies report only a single HFS, and examples of multiple HFSs within the same star-forming complex remain rare in the literature. Only a few studies have reported multiple parsec-scale HFSs in a single star-forming region \citep[e.g.,][]{Dewangan_2024, Maity_2025}. As described in Section~\ref{phy_env}, we identified four HFS candidates through visual inspection of the MIR images. Out of these only HFS-1 is associated with radio continuum emission. Existing studies show that clump c7 (or HFS-1/G35N) hosts EGO G35.20-0.74, driving molecular outflows \citep{Cyganowski_2008, Qiu_2013}. While the other HFSs does not show any signature of massive star formation, suggesting they are in the early stages of evolution.

Using the GRS $^{13}$CO(1--0) molecular line data, \citet{Dewangan_2017} proposed that CCC might be responsible for the birth of massive stars in the G35 cloud. Numerous MHD simulations have shown that CCC can lead to the formation of filamentary networks and, in some cases, HFSs \citep{Inoue_2013,Balfour_2015,Balfour_2017,Maity_2024}. These results are further supported by several observational evidence \citep{Furukawa_2009,Baug_2016,Dewangan_2017b,Hayashi_2018,Hitedoshi_2018,Maity_2022,Maity_2023}. Since G35N and G35S lie in the overlapping zone of two cloud components \citep[see Figure~7 in][]{Dewangan_2017}, it is possible that these HFSs, particularly HFS-1/G35N, formed through the CCC process. However, further analysis of this scenario lies beyond the scope of this study.

\subsection{Role of Magnetic fields}

B-fields play a crucial role in regulating the physical processes within molecular clouds and influencing the star formation activity \citep{Federath_2016}. Several observational and numerical studies have shown that, B-fields can influence gas dynamics by guiding the accretion flows along filaments towards the hubs, controlling fragmentation, and moderating the collapse of dense clumps \citep[e.g.,][]{Zhang_2014,Beuther_2018,Wang_2019, Cortes_2021, Wang_2022,Hwang_2022,Wang_2024, Khan_2024}.
However, the exact role of B-fields remains poorly understood, especially in massive star-forming regions \citep{Crutcher_2012}. In this context, we used the SOFIA/HAWC+ 154 $\mu$m observations to investigate the B-field morphology, strength, and energies toward two major clumps within the cloud (G35N and G35S).

\subsubsection{Magnetically Regulated Collapse in G35N}

G35N is a massive star-forming region associated with a massive ATLASGAL clump ($\sim$3100 $M_\odot$), an IR-dark HFS morphology, and compact radio emission. The POS B-field traced by both the SOFIA/HAWC+ and ACT datasets reveals an hourglass-shaped structure toward G35N (see Figure~\ref{fg4}). Such hourglass-shaped structures are also observed in other star-forming regions from clump to core scales \citep[e.g.,][]{Qiu_2014, Chuss_2019, Beltran_2019, Tahani_2023, Saha_2024, Beltran_2024, Law_2025}. However, we report the first detection of an hourglass-shaped B-field structure in a parsec-scale HFS.
Using the DCF method, we estimated a $B_{\rm POS}$ value of $\sim$600 $\pm$  200 $\mu$G (see Section~\ref{B-field strength}). The energy balance analysis indicates that gravity and B-fields contribute almost equally to the energy budget, followed by turbulence in G35N (Section~\ref{sec:energy}). This suggests that the clump is undergoing active gravitational collapse, regulated by B-fields. As mentioned earlier, \citet{Dewangan_2017} proposed a cloud-cloud collision (CCC) scenario toward G35 cloud. Using MHD simulation data, \citet{Maity_2024} showed that such collisions can amplify the B- field strengths from $\mu$G to mG levels \citep[see Figure 7 in][]{Maity_2024}, consistent with our estimated B-field values. In addition, \citet{Maity_2024} demonstrated the roles of turbulence, B-fields, and gravity in filamentary mass accretion onto the hub. Their results indicate that while turbulence is important for filament formation, B-fields and gravity dominate the later stages of gas collection, supporting our observational evidence for gravitational collapse regulated by B-fields.




%
Observationally, \citet{Qiu_2013} studied the B-field morphology in G35N using Submillimeter Array (SMA)  polarization data at $\sim$0.05 pc scales, showing a clear hourglass-shaped field toward its most massive core (see Figure~\ref{fig:afg2} in Appendix). Their mass-to-flux analysis confirmed that the core is magnetically supercritical, with gravitational forces dominating over magnetic pressure. Taken together, these results indicate that the hourglass morphology in G35N persists from clump ($\sim$5 pc) to core scales ($\sim$0.05 pc), providing strong evidence for magnetically regulated, active gravitational collapse across multiple spatial scales.
The preservation of the B-field structure across scales indicates that the field lines remain dynamically coupled to the gas, effectively regulating its motion and resisting distortion from turbulence and stellar feedback. In contrast, if the B-field morphology was not preserved, it would suggest that turbulence dominates over magnetic regulation, with the field structure being shaped primarily by local turbulent motions or gas dynamics. Studies at smaller spatial scales ($<$0.1 pc) toward G35N  have shown that gravity dominates over B-field, while turbulence plays a secondary role \citep{Qiu_2013, Hwang_2025}.

The signatures of massive star formation activity toward the G35N clump, suggests that the B-field is not enough to halt the gravitational collapse and subsqeuent star formation activity.
Some MHD simulations suggest that massive clumps are initially magnetically subcritical, with fields strong enough to resist gravitational collapse and channel material preferentially along the field lines \citep{Mouschovias_1976, Nakamura_2008}. As the clump accretes additional mass, it transitions to a magnetically supercritical state, where gravity overwhelms magnetic support and bends the field lines into the characteristic pinched (hourglass-like) morphology \citep{Gomez_2018, Beltran_2019} or spiral-like morphology \citep{Sanhueza_2021, Hwang_2022}. Crucially, this geometry is retained even in the supercritical phase as well. However, to confirm whether gas is indeed funneled along the B-field lines in G35N, high-resolution molecular line observations are essential. Such data would allow us to probe gas kinematics in relation to the field structure and establish the dynamical role of B-fields with greater certainty. Apart from channelling the accretion flows, the B-field lines also provide additional support which acts against gravity. This support delays the collapse of clumps and cores, allowing them to accumulate more mass before undergoing gravitational collapse \citep{Pillai_2015}. The buildup of these larger dense gas reservoirs ultimately favors the formation of massive stars. 

\subsubsection{Magnetic fields and stellar feedback in G35S}

In contrast to G35N, G35S hosts an evolved \htwo region (the Ap--21 Nebula), characterized by strong extended radio continuum emission \citep[see also][]{Dewangan_2017} and associated with two ATLASGAL clumps (c4 and c5). The B-field morphology inferred from the ACT 220 GHz data exhibits an hourglass-like structure, although with an axis of symmetry that differs from that observed in G35N. However, this morphology is not consistently seen in the B-field orientations derived from the SOFIA/HAWC+ 154 $\mu$m observations. The estimated B-field strength toward G35S is $\sim$850 $\pm$ 310 $\mu$G (see Section~\ref{B-field strength}). 
Energy balance analysis indicates that the B-field dominates over both gravity and turbulence in regulating the dynamics of the region (see Section~\ref{sec:energy}). Since B-field lines are effectively frozen into the molecular gas owing to the high conductivity of the medium, any changes in gas
dynamics directly alter the B-field structure of the cloud. Gas motions, such as compression,
expansion, or turbulence, can distort or realign the B-field lines. This coupling between gas dynamics and the B-field plays a crucial role in shaping the observed B-field morphology in
molecular clouds. 
To examine whether the expansion of the H\,{\sc ii} region could account for the high field strength, we estimated the ram pressure as $P_{\rm ram} = \rho v_{\rm exp}^2$, where $v_{\rm exp}$ is the expansion velocity of the H\,{\sc ii} region and $\rho$ is the mass density \citep[e.g.,][]{Hoang_2022}. The magnetic pressure is given by $P_{\rm B} = B^2/8\pi$, with $B$ as the POS B-field strength. Assuming typical expansion speed of 10 km s$^{-1}$ \citep{Faerber_2025}, and volume density of the clump. We find $P_{\rm ram}$ to be of order $10^{-7}$ dynes cm$^{-2}$, while $P_{\rm B}$ is of order $10^{-8}$ dynes cm$^{-2}$. Since the ram pressure exceeds the magnetic pressure, the expanding H\,{\sc ii} region can distort the field lines and enhance the local B-field strength.
In G35S, the expansion of the \htwo region compresses the surrounding neutral material due to the ram pressure. Since gas and dust are tightly coupled to the ions, this compression also strengthens and reshapes the B-field. As a result, feedback from the \htwo region not only alters the gas structure but also amplifies the B-field, making it the dominant force in the cloud's energy budget \citep[e.g.,][]{Devaraj_2021}.

\subsection{Interpreting Far-Infrared Polarization and important caveats}
\label{sec:extinction polarization}

The B-field orientations inferred from the SOFIA/HAWC+ at 154 $\mu$m and ACT at 1.3 mm show good agreement in G35N, whereas in G35S they are nearly perpendicular (see Figure~\ref{fg7}a). This contrasting behavior may arise for various reasons. First, the two wavelengths trace dust at different temperatures, which can lead to distinct B-field orientations in regions where the relative contributions of warm and cold dust differ.

Since G35S is an evolved \htwo region containing dust at a wide range of temperatures, the orientations inferred at 154 $\mu$m (a warm-dust tracer) may not align with those derived from the ACT 1.3 mm (cold-dust tracer) data. Second, because the orientations in G35S differ by  $\sim$90$^{\circ}$, the polarization observed at 154 $\mu$m may be dominated by extinction rather than thermal emission \citep[e.g.,][]{Novak_1997,Dowell_1997}. At 154 $\mu$m, polarization can arise from both dust emission and extinction depending on optical depth, whereas the polarization at 1.3 mm is almost entirely due to dust emission, since the emission at this wavelength is typically optically thin. Since the $\delta_\theta$ remains unchanged even without rotating the polarization vectors, the estimated field strength and energy balance conditions are unaffected.

We note that the polarization fractions measured toward the dense regions of G35N and G35S are low not only in the SOFIA/HAWC+ data but also in the ACT and Planck observations (see Figure~\ref{fig:afg4}). These low polarization fractions can increase susceptibility to systematic effects. Hence, a unique physical interpretation of the observed orientations cannot be established within the scope of the present work. In addition to temperature and optical depth effects, LOS variations in the B-field structure and grain alignment efficiency may also contribute to the differences in the inferred B-field directions between the SOFIA and ACT datasets toward the G35S region \citep{Fanciullo_2022}. A complex superposition of field geometries, dust properties, and alignment conditions could plausibly produce the differences between FIR and millimeter polarization without invoking a single mechanism. While a detailed LOS radiative transfer model is beyond the scope of this work, the observed multi-wavelength polarization in G35S likely reflects the combined influence of these effects rather than a simple, uniform B-field configuration.

Hence, assumptions that polarization at FIR wavelengths arises solely from emission must be treated with caution, and optically thin emission should be preferred (i.e., mm/ sub-mm wavelengths) for studying other massive star-forming regions. %
%
\section{Summary and Conclusions}
\label{sec:summary}

This paper focuses on multi-wavelength polarimetric observations from the SOFIA/HAWC+ at 154 $\mu$m and the ACT at 220 GHz to investigate the role of B-fields toward the G35N and G35S clumps in the G35 cloud. The key results of this study are as follows:

\begin{enumerate}
	\item Visual inspection of the {\it Spitzer} 8.0 $\mu$m and unWISE 12.0 $\mu$m images reveal at least four IR-dark HFSs (extent $\sim$2--3 pc) within the G35 cloud, including a prominent HFS toward G35N.
	
	\item The ACT 220 GHz polarization data show an hourglass POS B-field morphologies toward both the G35N and G35S sub-regions with distinct axes of symmetry.
	
	\item The SOFIA/HAWC+ 154 $\mu$m data reveal an hourglass-shaped B-field morphology toward HFS-1/G35N, whereas the B-field configuration toward G35S differs from that inferred from the ACT observations.
	
	\item The hourglass morphology identified at clump scales ($\sim$pc) in G35N aligns with the B-field structure previously reported at core scales ($\sim$0.05 pc), supporting a scenario of magnetically regulated collapse.
		
	\item The B-field strengths estimated using the DCF method from the SOFIA/HAWC+ 154 $\mu$m data are $\sim$$600 \pm 200~\mu$G toward G35N and $\sim$$850 \pm 310~\mu$G toward G35S.  
	
	\item The energy balance and the mass-to-flux ratio suggest that gravity and B-fields contribute comparably in G35N, while in G35S the gas dynamics are dominated by B-fields, followed by gravity and turbulence.
	
	\item The high B-field strength in G35S is likely a result of the expanding \htwo region, which compresses the surrounding material and amplifies the B-field strength.
	
	

	
\end{enumerate}

Overall, our results highlight the critical role of B-fields in regulating the star formation processes, with G35N showing signatures of a magnetically controlled collapse, while in G35S the field appears to be significantly shaped and amplified by the feedback from the expanding \htwo region.
\\

\section{Acknowledgments}
 We thank the anonymous referee for carefully reviewing the manuscript and providing valuable comments and suggestions, which helped improve the scientific content of this paper. The research work at Physical Research Laboratory is funded by the Department of Space, Government of India.  We acknowledge the support from Ajman University through the Internal Research Grant No. DRGS Ref. 2025-IRG-CHS-6. This work was also supported by the Deanship of Scientific Research, Vice Presidency for Graduate Studies and Scientific Research, King Faisal University, Saudi Arabia, under Proposal Number KFU253355. PS was partially supported by a Grant-in-Aid for Scientific Research (KAKENHI Number JP22H01271 and JP23H01221) of JSPS. We are grateful to Dr.~N.~K Bhadari for kindly providing information on ACT data, which greatly aided this work. The MeerKAT telescope is operated by the South African Radio Astronomy Observatory, which is a facility of the National Research Foundation, an agency of the Department of Science and Innovation. The National Radio Astronomy Observatory is a facility of the National Science Foundation operated under cooperative agreement by Associated Universities, Inc. This research is based on observations made with the NASA/DLR Stratospheric Observatory for Infrared Astronomy (SOFIA). SOFIA is jointly operated by the Universities Space Research Association, Inc. (USRA), under NASA contract NNA17BF53C, and the Deutsches SOFIA Institut (DSI) under DLR contract 50 OK 0901 to the University of Stuttgart. This work is based, in part, on observations made with the {\it Spitzer} Space Telescope, which is operated by the Jet Propulsion Laboratory, California Institute of Technology under a contract with NASA. This research made use of Astropy, a community-developed core Python package for Astronomy \citep{astropy:2013,astropy:2018,astropy:2022}.

\begin{figure*}
	\centering
	\includegraphics[width= 19 cm]{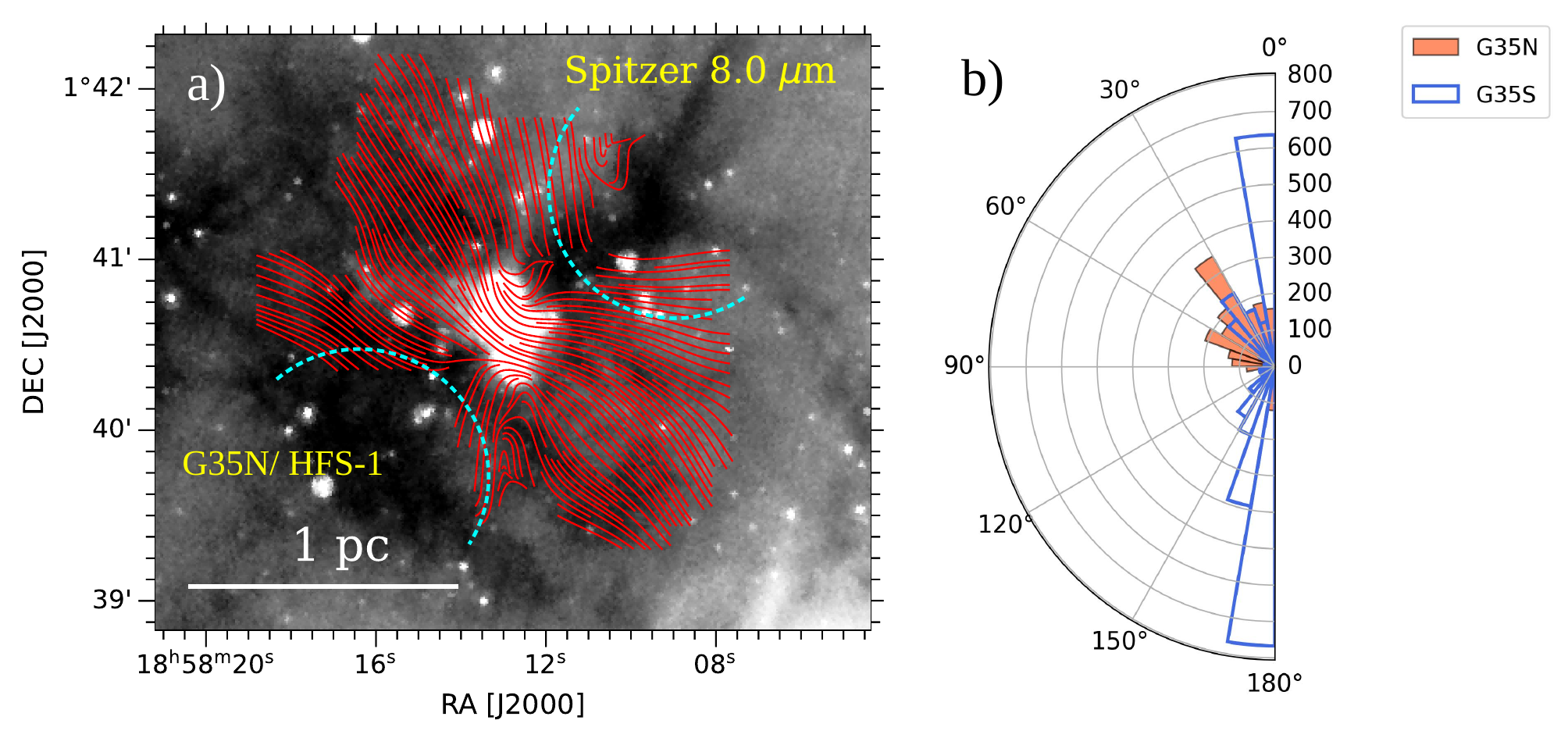}
	\caption{a) {\it Spitzer} 8.0 $\mu$m image of HFS-1/G35N overlaid with the streamlines showing the B-field directions estimated using the SOFIA/HAWC+ 154 $\mu$m polarization data. The dotted curves highlight the hourglass-like structure of the B-fields. The scale bar is the same as shown in Figure~\ref{fg3}a. b) The panel depicts the histogram of B-field position angles from the SOFIA/HAWC+ 154 $\mu$m toward the G35N (in orange), and G35S (in blue). }
	\label{fg4}
\end{figure*}

\begin{figure*}
	\centering
	\includegraphics[width= 19 cm]{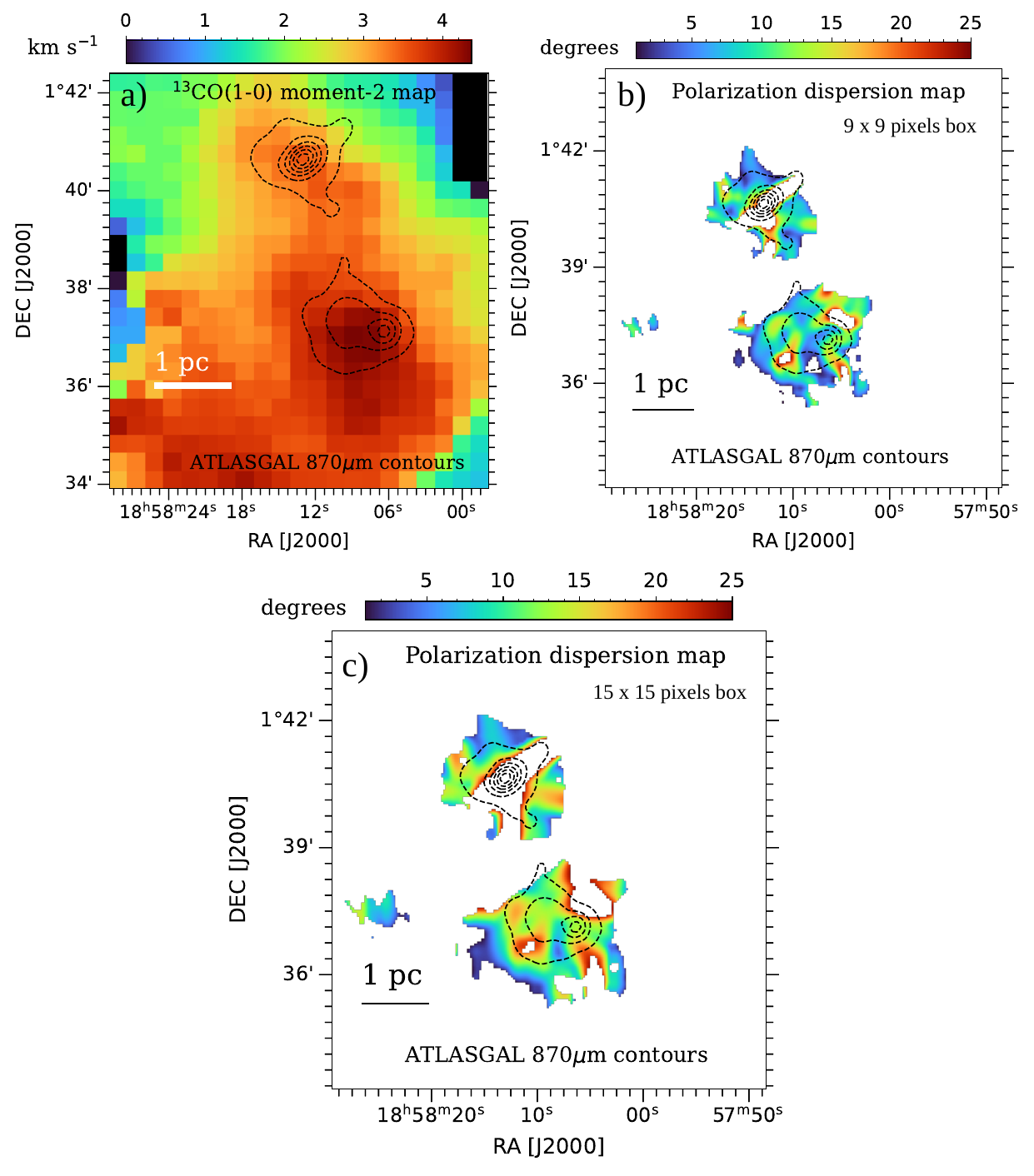}
	\caption{a) The panel shows the GRS $^{13}$CO(1--0) moment-2 map toward an area shown in Figure~\ref{fg3}a.b) The polarization dispersion ($\delta\theta$) map toward the G35 cloud using 9 $\times$ 9 pixels sliding box. c) $\delta\theta$ map toward G35 cloud using 15 $\times$ 15 pixel sliding box.
	The dashed contours represent the ATLASGAL 870$\mu$m continuum emission. The contour levels are 1, 3.8, 6.5, 9.3, and 12 Jy beam$^{-1}$. The scale bar is the same as shown in Figure~\ref{fg3}a.}
	\label{fg5}
\end{figure*}

\begin{figure*}
	\centering
	\includegraphics[width= 19 cm]{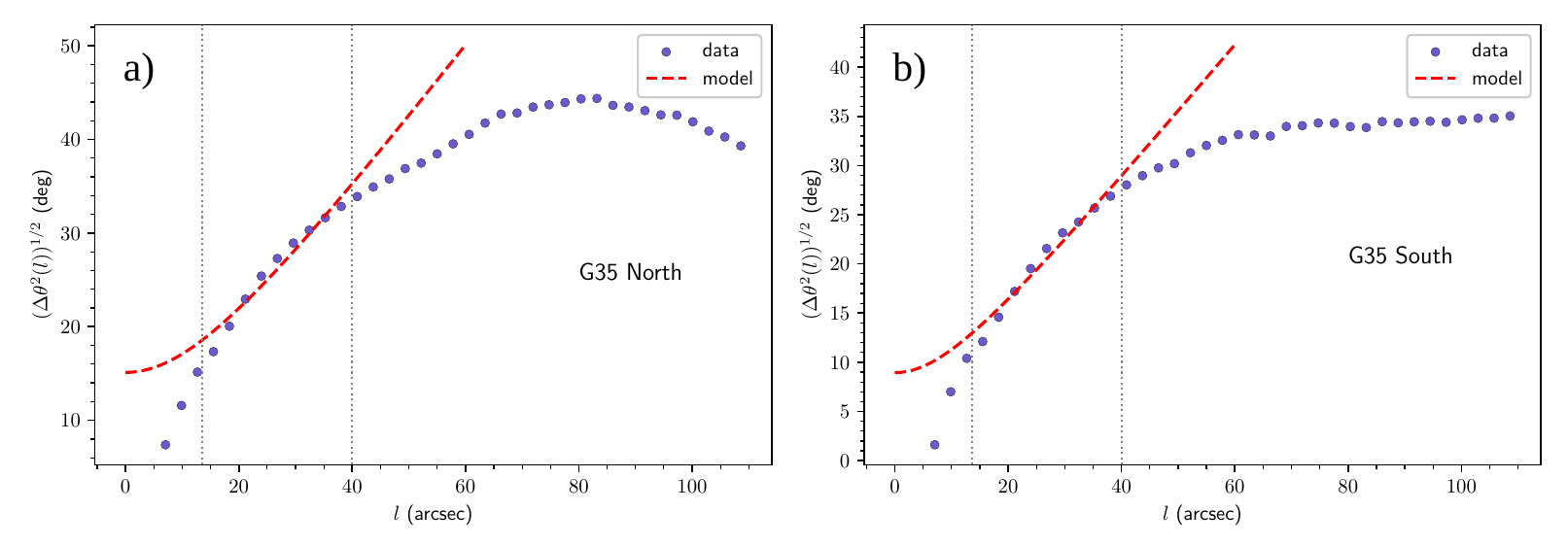}
	\caption{Panels (a) and (b) present the structure function ($D^{1/2}_\theta(\ell)$) for the G35N and G35S sub-regions of the cloud. The red-dashed line shows the linear fit of Equation~\ref{eq:quadratic_func}, while the grey vertical dashed lines mark the lower and upper limits used for the fit.}
	\label{fg6}
\end{figure*}

\begin{figure*}
	\centering
	\includegraphics[width= 19 cm]{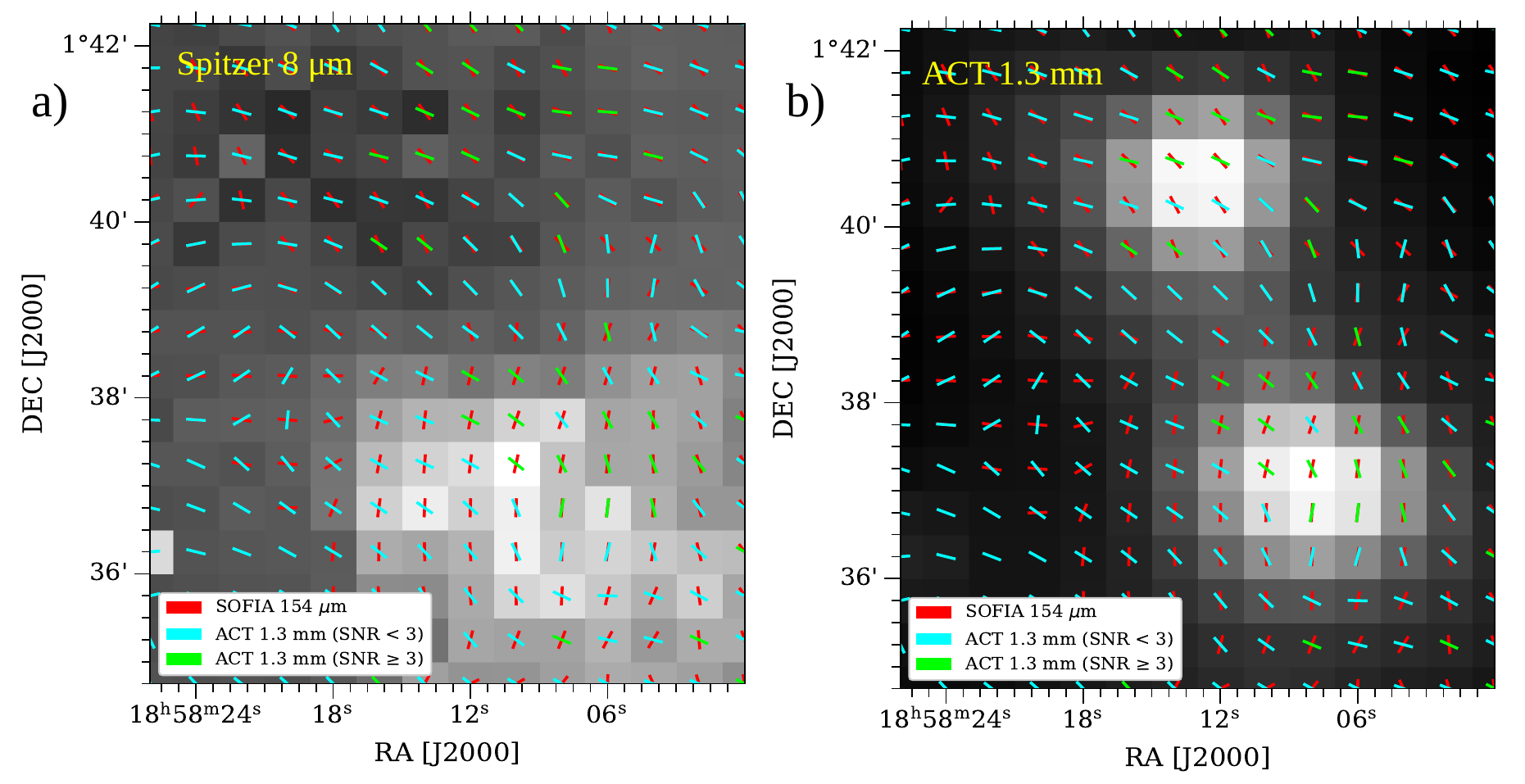}
	\caption{a) {\it Spitzer} 8.0 $\mu$m image (in the inverse hyperbolic sine scale and re-gridded to same pixel scale as the ACT 220 GHz data) overlaid with B-field orientations from the SOFIA/HAWC+ 154 $\mu$m (in red) and ACT 220 GHz (in cyan). b) Same as Figure~\ref{fg7}a, the back ground image is the ACT 220 GHz continuum map. The cyan and green segments represent polarization vectors with $p/\sigma_p<3$ and $p/\sigma_p\geq3$, respectively.} 
	\label{fg7}
\end{figure*}


\bibliographystyle{aasjournal}
\bibliography{bibfile}

\appendix
\restartappendixnumbering

\section{Physical Parameters of the ATLASGAL clumps toward G35 cloud}
\label{s1sec:d} 
\begin{figure*}[ht!]
	\centering
	\includegraphics[width=17 cm]{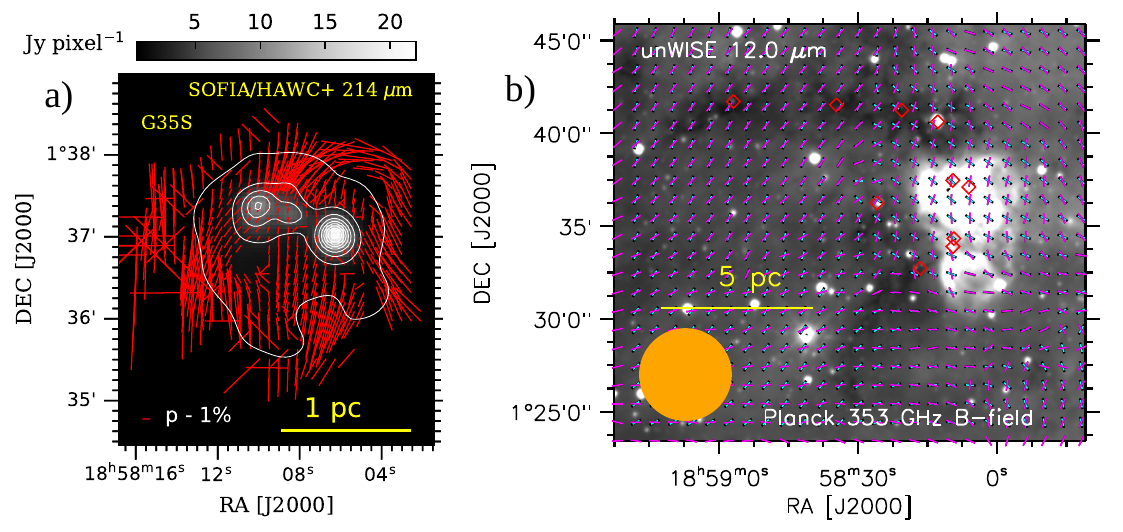}
	\caption{a) Overlay of segments showing POS B-field orientations from SOFIA/HAWC+ 214 $\mu$m observations on the SOFIA/HAWC+ 214 $\mu$m continuum map (see white box in Figure~\ref{fg3}b). The segments satisfy the following criteria: $p/\sigma_p>$3, $I/\sigma_I>$10, and $0.6<p<30$ $\%$. The lengths of the segments are proportional to the degree of polarization. A reference for 1 $\%$ degree of polarization is shown in the bottom left corner of the image. The contour levels are 1, 3.8, 6.5, 9.3, and 12 Jy beam$^{-1}$. A scale bar of 1 pc at a distance of 2.19 kpc is shown.
	b) Overlay of segments showing the B-field orientations inferred from the {\it Planck} 353 GHz polarization data (magneta) and from the ACT 1.3 mm polarization data (cyan) on the unWISE 12.0 $\mu$m map (in inverse hyperbolic sine scale). 
	A filled circle shows the beam size of 5$'$. A scale bar of 5 pc at a distance of 2.19 kpc is shown. Diamonds highlight the positions of the ATLASGAL clumps (see Table~\ref{tab2}).}
	\label{fig:afg3}
\end{figure*}

This section presents the ATLASGAL clumps \citep{Urquhart_2017} distributed in the target area, and their physical properties are listed in Table~\ref{tab2} (see also diamonds in Figure~\ref{fig:afg3}). 
\begin{table*}[ht!]
	\renewcommand{\arraystretch}{1.1} 
	\centering
	\begin{tabular}{ccccccccccc}
		\hline
		ID & RA &   Dec &              Clump &  $F_{\rm peak}$ &   $F_{\rm int}$ &  $V_{\rm lsr}$ & $d$ &   Radius  &  $T_{\rm dust}$ &  $M_{\rm clump}$ \\
		& (hh:mm:ss)& (dd:mm:ss) & & (Jy beam$^{-1}) $& (Jy) & (km s$^{-1}$) & (kpc) & (pc) & (K) & ($M_{\odot}$)\\
		\hline
		1 & 18:58:16.5391 & +1:32:43.684 & AGAL035.087-00.816 &   0.44 &   2.24 &  34.2 &  2.19 & 0.1 &   13.2 &  116 \\
		2 & 18:58:09.5015 & +1:33:54.022 & AGAL035.091-00.781 &   0.43 &   2.66 &  34.7 &  2.19 & 0.1 &   18.3 &   81 \\
		3 & 18:58:09.3047 & +1:34:19.812 & AGAL035.097-00.777 &   0.49 &   3.65 &  34.4 &  2.19 & 0.13 &   17.8 &  115 \\
		4 & 18:58:06.0920 & +1:37:06.144 & AGAL035.132-00.744 &   7.85 &  39.19 &  35.0 &  2.19 & 0.43 &   19.4 & 1086 \\
		5 & 18:58:09.5434 & +1:37:28.138 & AGAL035.144-00.754 &   4.52 &  15.10 &  32.9 &  2.19 & 0.12 &   21.7 &  355 \\
		6 & 18:58:25.9209 & +1:36:14.709 & AGAL035.157-00.824 &   0.74 &   3.23 &  31.8 &  2.19 & 0.15 &   15.2 & 1310 \\
		7 & 18:58:12.7898 & +1:40:37.599 & AGAL035.197-00.742 &  12.35 & 132.07 &  33.9 &  2.19 & 1.64 &    - & 3097 \\
		8 & 18:58:20.5475 & +1:41:15.009 & AGAL035.221-00.766 &   0.77 &   7.91 &  35.4 &  2.19 & 0.34 &   14.7 &  340 \\
		9 & 18:58:34.6265 & +1:41:32.091 & AGAL035.252-00.816 &   0.48 &   3.85 &  36.6 &  2.19 & 0.1 &   11.8 &  243 \\
		10 & 18:58:56.8622 & +1:41:43.028 & AGAL035.297-00.897 &   1.01 &   4.51 &  37.5 &  2.19 & 0.15 &   15.4 &  179 \\
		\hline
	\end{tabular}
	\label{tab2}
	\caption{Physical parameters of the ATLASGAL clumps toward the G35 cloud.}
\end{table*}

\section{Magnetic field morphology toward G35N from SMA}
\label{sec:app2wx}
\begin{figure*}[ht!]
	\centering
	\includegraphics[width=18cm]{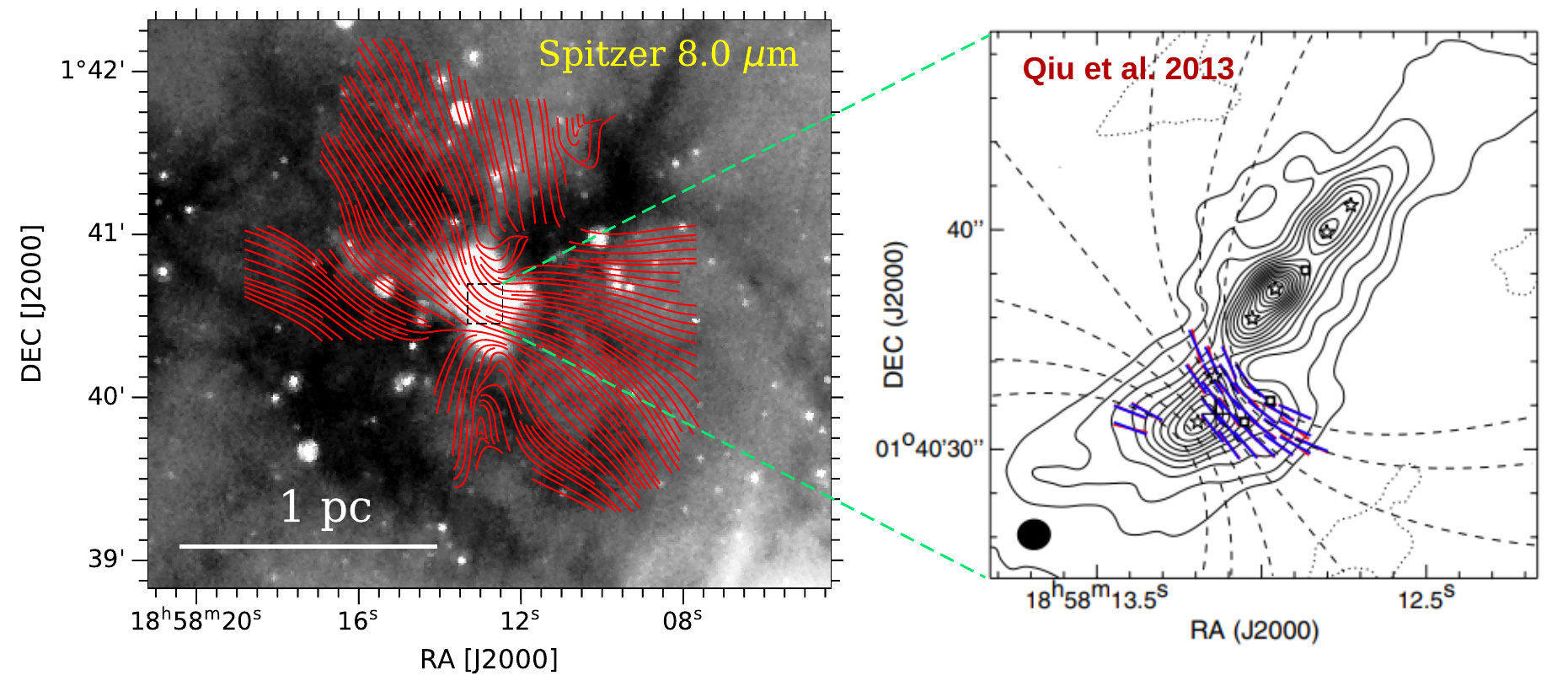}
	\caption{a) The panel shows the {\it Spitzer} 8.0 $\mu$m image of HFS-1/G35N overlaid with the streamlines showing the B-field directions estimated using the SOFIA/HAWC+ 154 $\mu$m polarization data. A scale bar corresponds to 1 pc at a distance of 2.19 kpc  (see also Figure~\ref{fg4}a). b) B-field morphology toward black dashed box in Figure~\ref{fig:afg2}a shown as red segments taken from \citet{Qiu_2013} toward the central region of HFS-1. \textcopyright{} AAS. Reproduced with permission. The black contours are SMA 880 $\mu$m continuum emission. The blue segments show the tangent directions of the fitted parabola at the locations of B-field measurements.}
	\label{fig:afg2}
\end{figure*}

Using the SMA 880 $\mu$m polarization observations (beam size: $\sim$1\rlap.{$''$}0 $\times$ $\sim$0\rlap.{$''$}6), the B-field morphology toward G35N was previously investigated by \citet{Qiu_2013} . Figure~\ref{fig:afg2}b shows the observed B-field orientations \citep[from][]{Qiu_2013} toward the most massive core in G35N (red segments), along with the best-fit parabolic curves (blue segments; dashed black lines).
	
\newpage
\section{Polarization fraction toward the G35 cloud from Planck 353 GHz and ACT 220 GHz}
\label{sec:app4wx}

\begin{figure*}[ht!]
	\centering
	\includegraphics[width=19cm]{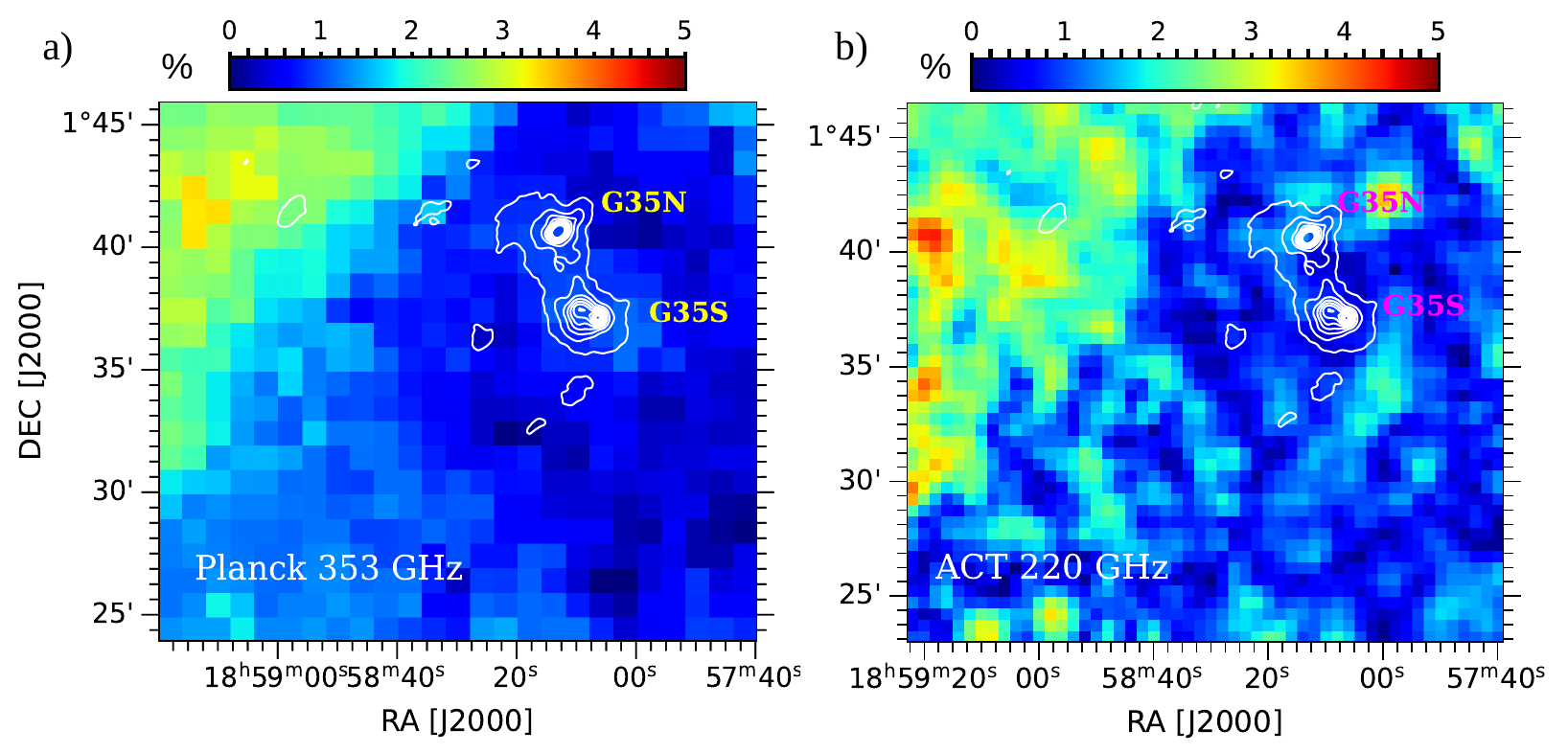}
	\caption{a) The panel shows the polarization fraction of Planck 353 GHz data toward G35 cloud shown in Figure~\ref{fg1}a overlaid with ATLASGAL 870 $\mu$m contours. The contour levels are same as Figure~\ref{fg1}a. b) Same as Figure~\ref{fig:afg4} but for ACT 220 GHz data. }
	\label{fig:afg4}
\end{figure*}

The $p$ toward the G35.20-0.74 star-forming complex for Planck 353 GHz and ACT 220 GHz observations is shown in Figures~\ref{fig:afg4}a and \ref{fig:afg4}b, respectively. The $p$ values toward G35N and G35S are around $\sim$1$\%$ in the Planck 353 GHz data, while the ACT 220 GHz observations show slightly higher values of $\sim$2$\%$.

\end{document}